\documentclass[12pt]{iopart}

\usepackage{subfigure}

\usepackage{graphicx}
\usepackage{mathrsfs}
\usepackage[british]{babel}

\expandafter\let\csname equation*\endcsname\relax 
\expandafter\let\csname endequation*\endcsname\relax 

\usepackage{amsbsy,amssymb,amsmath,bm}
\usepackage{graphicx,color,epsfig,rotate}

\catcode`@=11 
\renewcommand\tableofcontents{%
  \section*{\contentsname}%
  \@starttoc{toc}%
}
\catcode`@=12

\newcommand{\Ket}[1]{\left|#1  \right>}

\newcommand{\beq}{\begin{equation}}
\newcommand{\eeq}{\end{equation}}

\usepackage{multirow}

\begin{document}

\title[Nonequilibrium quantum dynamics and transport: from integrability to MBL]{Nonequilibrium  quantum dynamics and transport: from integrability to many-body localization}

\author{Romain Vasseur and Joel E. Moore}
\address{${}^1$Department of Physics, University of California, Berkeley, California 94720, USA}
\address{${}^2$Materials Sciences Division, Lawrence Berkeley National Laboratory, Berkeley, California 94720, USA}

\eads{\mailto{rvasseur@berkeley.edu}, 
      \mailto{jemoore@berkeley.edu}}

\date{\today}

\begin{abstract} We review the non-equilibrium dynamics of many-body quantum systems after a quantum quench with spatial inhomogeneities, either in the Hamiltonian or in the initial state. We focus on integrable and many-body localized systems that fail to self-thermalize in isolation and for which the standard hydrodynamical picture breaks down. The emphasis is on universal dynamics, non-equilibrium steady states and new dynamical phases of matter, and on phase transitions far from thermal equilibrium. We describe how the infinite number of conservation laws of integrable and many-body localized systems lead to complex non-equilibrium states beyond the traditional dogma of statistical mechanics.

\end{abstract}

\newpage

\tableofcontents

\section{Introduction} 

Quantum systems of many particles far from equilibrium pose notable challenges for theory as they are not susceptible to the general principles and methods that make many equilibrium systems tractable.  Until quite recently, the difficult questions about what sorts of transport and dynamical processes occurred away from equilibrium were postponed.  These questions are now central to the progress of theoretical physics: new experiments probe quantum coherent dynamics in regimes where conventional semiclassical approaches are clearly inadequate, and new theoretical concepts indicate that quantum effects on many-particle dynamics are richer and more interesting than previously suspected.

The two categories of non-equilibrium problems that we address in this article involve two different connections between integrability and spatial inhomogeneity, i.e., breaking of spatial translation invariance.   Conventional integrable systems, such as the Heisenberg antiferromagnet or Lieb-Liniger gas, are translation-invariant problems in one spatial dimension.  However, inhomogeneous initial conditions arise in a variety of physical problems, particularly those related to transport of particles or energy.  Linear response theory gives an approach to {\it near-equilibrium} transport based on dynamical correlation functions at equilibrium, but far-from-equilibrium transport requires a genuinely non-equilibrium framework, and probably the most natural far-from-equilibrium transport problem concerns the current that develops between two reservoirs at different temperatures or chemical potentials.  The first part of this review is devoted to transport in integrable models induced by inhomogeneous initial states, where special properties of certain integrable models turn out to be very powerful in constraining particle and energy flow.

The second category of problems discussed here is related to a different kind of integrability.  It is now understood that the many-body localized (MBL) phase that generalizes single-particle Anderson localization to strong interactions has, like the Anderson case, an infinite number of {\it local} conserved quantities~\cite{PhysRevLett.111.127201,PhysRevB.90.174202,VoskAltmanPRL13,2014arXiv1403.7837I}.  Here ``local'' means that the conserved operator is exponentially localized in real space, while the conserved quantities in normal integrable models are translation-invariant sums of local quantities.  The focus here is on universal behavior in the many-body localized phase and possibly at the localization transition, and specifically on how such behavior can differ from the Anderson localized case, in part because the conserved quantities interact with each other in the Hamiltonian in the case of MBL. As we will describe in more detail below, some of the standard questions about a quantum system need to be modified in the MBL context because an MBL system does not generally thermalize, but there are nevertheless important and robust properties in the long-time dynamics that one can hope will be observed in experiment now that MBL systems are being created~\cite{PhysRevLett.114.083002,Schreiber842,2015arXiv150807026S,2015arXiv150900478B}.

There are several features common to the two categories for the reader to keep in mind.  In both cases, conserved quantities provide the key to understanding many important properties.  Integrable models can be understood as having unconventional ``hydrodynamics'': conventional hydrodynamics results from decay of all but a few slow modes associated with standard conservation laws of particle number, energy, and momentum.  In an integrable model of either conventional or MBL type, the simple fact that there is an infinite number of conservation laws suggests that our standard intuition about hydrodynamical behavior needs to be modified.  It is sometimes useful to think of conventional integrable models starting from free (i.e., quadratic) models such as free fermions or the $XX$ chain, and to think of the many-body localized phase starting from the single-particle Anderson localized phase, and we will frequently use free systems as examples.  The interacting system we discuss in greatest detail is the pure or random XXZ chain, equivalent to spinless fermions with nearest-neighbor interactions.  In both cases the challenge is to obtain non-trivial results on the new physics induced by interactions.

It is also crucial to remark that both integrable and MBL systems are special -- and interesting! -- from the point of view of non-equilibrium quantum dynamics, because they fail to thermalize. In this review, we will be interested in the dynamics of an isolated,  interacting many-body quantum system after a generic global quench, where an extensive amount of energy is injected into the system at time $t=0$. If the system is generic ({\it i.e.} non-integrable), the natural expectation is that it should eventually go back to thermal equilibrium, and that statistical mechanics should naturally emerge from quantum mechanics: this is the scenario of thermalization. The way this happens in an isolated quantum system is quite nontrivial since the dynamics is unitary and no information is loss during time evolution, so that in principle, the system always ``remembers'' its initial state. However, within such thermalizing systems, the local memory of the initial condition is essentially lost (by decoherence) and the system acts as its own heat bath to reach thermal equilibrium with an effective temperature that will depend only on the energy density of the initial state. In other words, at very long times the memory of the initial condition is hidden in global degrees of freedom that cannot be accessed with physical, local observables, so that the system essentially forgets about its initial state despite the unitary dynamics. This leads to an effective, statistical mechanics description of local observables in terms of a few  parameters like temperature, chemical potential {\it etc.} -- one per extensive conserved quantity in the system. 

Interestingly, not all interacting many-body quantum systems thermalize. As we will see in the following, integrable systems fail to reach thermal equilibrium, leading to new non-equilibrium steady states and to singular transport properties. However, integrable systems are in general highly fine-tuned, and generic perturbations will break integrability. Whereas signatures of integrability can persist for a very long time corresponding to a ``pre-thermalization'' regime (see {\it e.g.}~\cite{2016arXiv160309385L} in this special issue), generic quantum systems tuned away from special integrable points are commonly expected to eventually reach thermal equilibrium. MBL systems provide a more robust version of integrability that emerge naturally in strongly disordered systems, leading to a variety of new quantum phases and dynamical phase transitions far away from equilibrium. In the following, we describe how the conserved quantities of integrable and MBL systems lead to various non-equilibrium states beyond the traditional tenets of statistical mechanics.

\section{Quantum transport in integrable systems}

We start by summarizing some basic expectations regarding conventional (linear-response) transport in integrable systems.  Linear response transport coefficients are expressed via the Kubo formula as integrals over equilibrium dynamical correlation functions.  It is worth thinking for a moment about what these formulas mean in a system that does not thermalize or thermalizes unconventionally (e.g., to a generalized Gibbs ensemble~\cite{2016arXiv160403990V} that includes more conservation laws than just energy).

In a system with global charge conservation, the Kubo formula for electrical conductivity at zero frequency in a one-dimensional system of length $L$ reads
\begin{equation}
\sigma(\omega=0)=\frac{1}{2LT}\int_0^{+\infty}\Re\langle J(t)J(0)\rangle_T\,dt,
\label{Kubo}
\end{equation}
where the angle brackets indicate averaging over the thermal ensemble at temperature $T$.  The spirit of the Kubo formula (\ref{Kubo}) is that the return to equilibrium from a perturbation (an applied electric field), which is described by $\sigma$, should be essentially similar to the return to equilibrium from a spontaneous fluctuation.  For an integrable system, we shall take (\ref{Kubo}) as the definition of the DC conductivity, realizing that at a minimum the definition should be generalized to include other conserved quantities in the ensemble.  

Even then $\sigma$ need not be well-defined as written; a simple example is a system of one species of charged particles moving in the continuum, which will have a divergent zero-frequency conductivity as the current cannot relax.  More precisely, if current is proportional to a conserved quantity such as momentum, the time integral in (\ref{Kubo}) diverges.  We can measure the degree of divergence by generalizing conductivity to finite frequency and defining the Drude weight $D$:
\beq
\sigma(\omega) = D(T) \delta(\omega) + \ldots.
\eeq
Integrable systems often have dissipationless transport with a nonzero ``Drude weight'' $D$ in their frequency-dependent conductivity~\cite{PhysRevB.55.11029,PhysRevB.77.245131}.
This is naturally connected to a limited form of thermalization: if thermalization occurs then the current relaxes and the Drude weight is zero, while if thermalization does not occur then the Drude weight can still in principle be zero if the current operator is not sufficient to probe the failure of thermalization.

Considerable progress has been made in recent years on understanding the Drude weight in integrable models, and particularly in the XXZ spin chain:
\begin{equation}
H=\sum_{i=1}^{L-1}\left(S_i^xS_{i+1}^x+S_i^yS_{i+1}^y+\Delta S_i^zS_{i+1}^z+ h S_i^z\right).
\label{XXZ}
\end{equation}
We summarize the expected behavior of integrable and non-integrable models as a prelude to our study of far-from-equilibrium properties in the following section.  The Drude weight for a current is bounded below by a sum over conserved quantities that have a nonzero projection on to the current, via the \emph{Mazur inequality}~\cite{Mazur1969533,PhysRevB.55.11029}
\begin{equation}
	\label{quenches_MazurIneq}
	\lim_{t\to\infty}(\langle P(t)P(0)\rangle - \langle P\rangle^{2}) \geq \sum_{\alpha}\frac{\langle P Q_{\alpha}\rangle^{2}}{\langle Q_{\alpha}Q_{\alpha}\rangle},
\end{equation}
where $Q_{\alpha}$ are independent local conserved quantities~\cite{SUZUKI1971277}.  In the XXZ model at zero magnetic field, the conventional conserved quantities give zero contribution to the Drude weight by symmetry~\cite{sirker:2010}, but a new set of conserved quantities~\cite{prosenxxz} do contribute.  These new quantities were discovered from consideration of far-from-equilibrium steady states of the type discussed in the following section.  At least at high temperatures and some values of anisotropy $\Delta$, these new quantities appear to saturate the numerical value of the Drude weight~\cite{karraschdrude} obtained from improved time-dependent matrix-product-state simulations.

The same numerical methods suggest that when an integrability-breaking staggered field is added to the XXZ model (but still without disorder), then the conductivity becomes finite in the region where this perturbation is irrelevant~\cite{huangkarrasch}; the values are consistent with theoretical predictions from a bosonization approach~\cite{sirker:2010,PhysRevB.83.035115} and may indicate thermalization although that would require longer simulation times and consideration of additional quantities beyond the current.  This demonstrates an important point that is sometimes obscured in the literature.  It is commonplace to think that an irrelevant operator means one that is unimportant at low frequencies and hence long times, but this requires some care in the context of integrability breaking.  It is true that as the temperature is lowered, an irrelevant integrability-breaking perturbation will take longer to have its effect.  However, at any nonzero initial temperature, the ultimate long-time behavior of the system is dominated by the irrelevant perturbation.  In other words, the expectation that integrability-breaking perturbations will ultimately induce thermalization means that they can be ``dangerously irrelevant'' perturbations in the long-time dynamics, even if they are ordinary irrelevant perturbations from the point of view of thermodynamic properties.  Integrability breaking has similar far-reaching effects in the strongly nonequilibrium situation we now consider.

\section{Global quenches and non-equilibrium steady states}

\begin{figure}
\begin{center}
\includegraphics[width = 0.85\columnwidth]{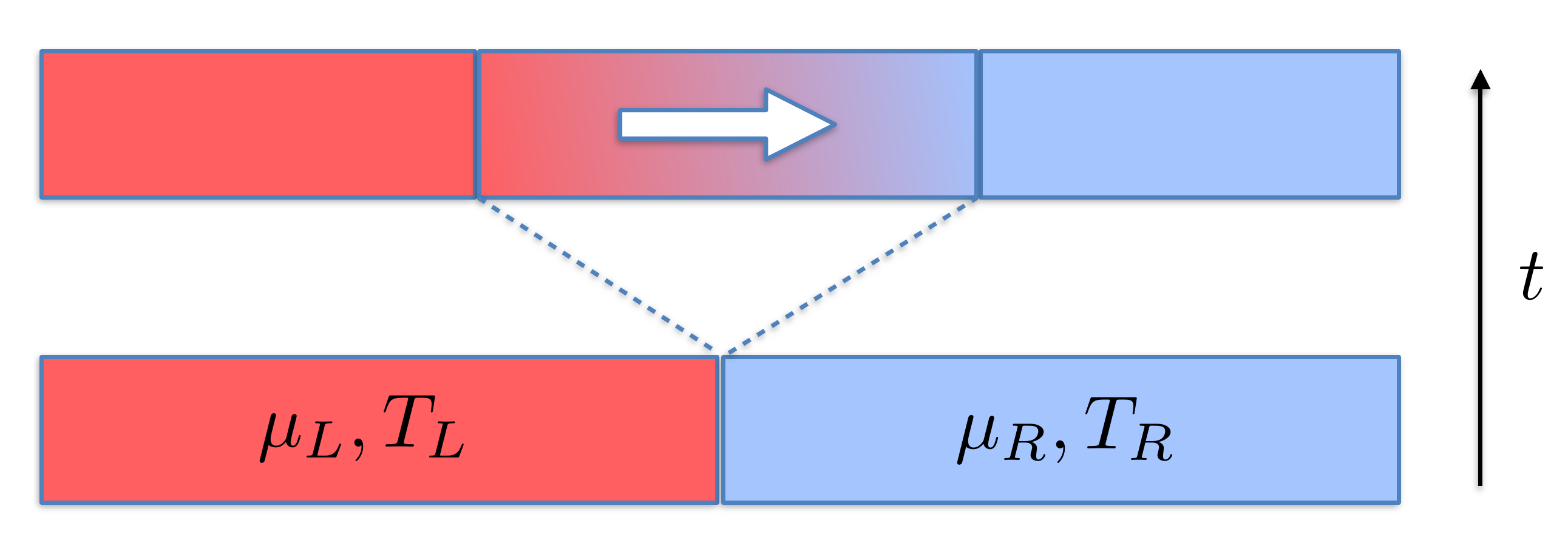}
\vspace{-.2in}
\end{center}
\caption{A geometry for far-from-equilibrium transport: two half-line reservoirs are prepared at one temperature or chemical potential for $x<0$ and a different temperature or chemical potential for $x>0$.  At $t=0$ the left and right reservoirs are connected in such a way that the final system is translation-invariant.}
\label{Fig1}
\end{figure}

\subsection{Non-interacting systems and Stefan-Boltzmann law}

We now review how free systems, such as the XXZ model at $\Delta = 0$, behave outside the linear-response regime.  That requires consideration of a genuinely non-equilibrium situation, and we focus on the popular~\cite{sotiriadiscardy,bernarddoyon,karraschilanmoore,BernardDoyonReview,2016arXiv160307765B} geometry shown in Figure~\ref{Fig1}.  Two reservoirs, each at equilibrium, are connected at $t=0$, and the subsequent evolution under a constant translation-invariant Hamiltonian is observed.  We focus for now on the case of a temperature difference at zero chemical potential~\footnote{We use the term ``chemical potential'' rather than ``magnetic field'' because all the evolution takes place in zero magnetic field and the role of the chemical potential is to prepare the initial equilibria.  If we had genuinely different magnetic fields on the two half-lines then the Hamiltonian would not be translation-invariant)}.  This problem is one of the simplest non-translation-invariant ``quantum quenches'' if one views the original system as prepared with a Hamiltonian that is cut at $x=0$; then the quench is simply the restoration of the link at $x=0$.  We note that quantum quenches have been studied a great deal over the past ten years (see {\it e.g}~\cite{castin_dum,shlyapnikov_expansion,stringari_expansion,gangardt_pustilnik,fleischauer,caux_konik,PhysRevLett.98.050405,gritsev_demler,caux_essler,iyer_andrei,bloch_expansion,gangardtbose,PhysRevB.89.075139} for related quenches addressing the expansion of a wave-packet into the vacuum), and we refer the interested reader to the other reviews in this special issue for a more exhaustive account of quantum quenches in setups different than Fig.~\ref{Fig1} (see {\it e.g}~\cite{2016arXiv160308628D,2016arXiv160306452E,2016arXiv160304689C,2016arXiv160304252C,2016arXiv160302889C}).

\begin{figure}

\begin{center}
\includegraphics[width = 0.85\columnwidth]{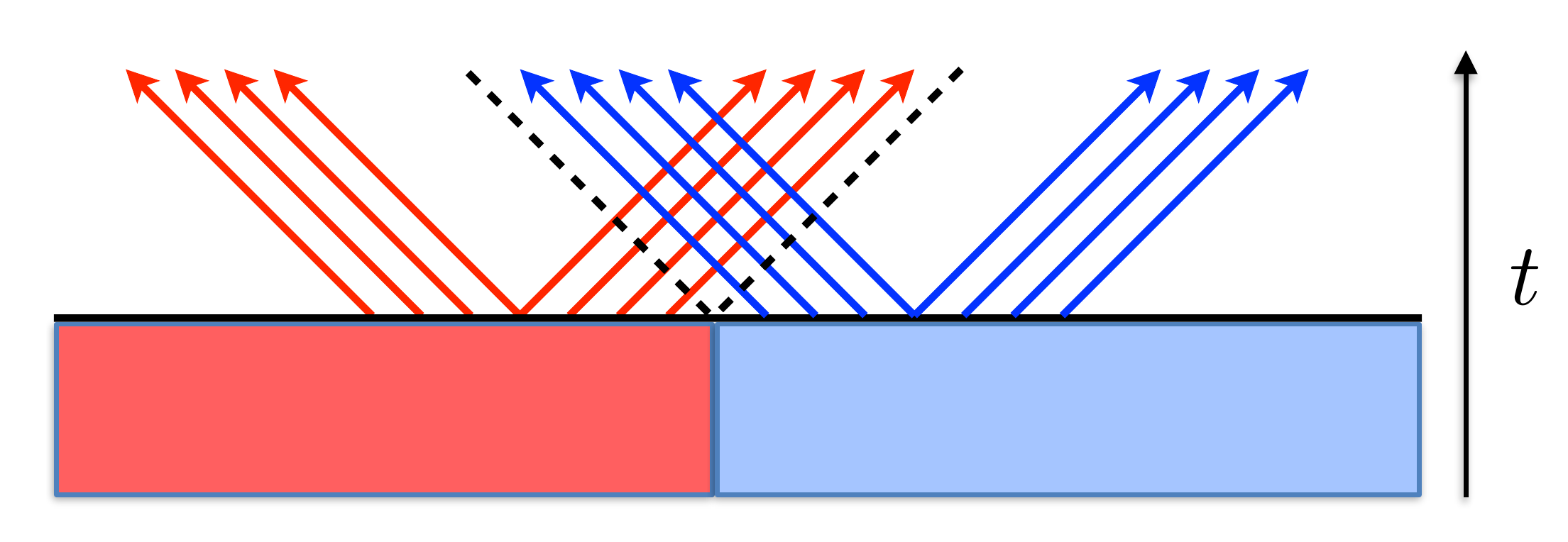}
\vspace{-.2in}
\end{center}

\caption{Absence of local equilibrium in the free (or conformal) case: excitations are radiated at one temperature from the left half-line, and a different temperature from the right half-line.  The absence of interactions means that points in the central region do not have a well-defined temperature, because a different temperature is required to characterize left- and right-movers.}
\label{Fig2}
\end{figure}

The key idea is shown in Fig.~\ref{Fig2}: as time increases, there is a larger and larger central region over which the local state of system can be described as having a distribution of right-movers equivalent to the distribution of right-movers in the thermal equilibrium that described the left half-line, and likewise a distribution of left-movers equivalent to the distribution of left-movers in the right half-line.  For the XX model, this leads to simple calculations for transport in the steady state that can be confirmed by numerical simulations~\cite{karraschilanmoore}.  A more universal setting for this idea is to consider one-dimensional quantum systems that are described by conformal field theories: the thermal current at late times at a point in the central region (for example, the origin) is found to be~\cite{sotiriadiscardy,bernarddoyon}
\begin{equation}
j^\infty_E = {\pi c \over 12} {k_B^2 \over \hbar} {\left(T_L^2 - T_R^2\right)},
\label{cft1d}
\end{equation}
where $c$ is the central charge and $k_B$ is Boltzmann's constant.

An intuitive understanding of (\ref{cft1d}) is that the energy current is given by the difference in radiated power, described by the Stefan-Boltzmann law, from the two reservoirs:
\begin{equation}
j^\infty_E(T_L,T_R) = f(T_L) - f(T_R),
\label{ffunction}
\end{equation}
where the Stefan-Boltzmann function $f$ goes as $T^2$ in two spacetime dimensions for the same reasons it goes as $T^4$ in four dimensions.  One way to derive the very universal form of the formula for $f$, which depends only on central charge, is to realize that the thermal conductivity (the derivative of $f$) multiplies the specific heat, which has the central charge and an inverse power of the conformal velocity~\cite{affleckanomaly}, by the velocity.

The same notion of a Stefan-Boltzmann function, albeit of a more complicated form, applies in a general free theory even without conformal invariance, and applies equally well to charge transport in such theories.  As an explicit example, consider the XX model in fermionic variables with $\epsilon_k$ being the energy of the excitation at wavevector $k$.
We can use the Landauer approach~\cite{landauer} to compute the thermal current: this amounts to computing the right-moving energy current from a lead at temperature $T_L$ and subtracting the left-moving energy current from a lead at $T_R$.  Using $k$ for momentum, we have that the total energy current (units of energy per time) is
  \begin{equation}
  J_E = J_E^R - J_E^L = \int_0^{\pi} \,{dk \over 2 \pi} \,\left[f_{T_L}( \epsilon_k) -f_{T_R}(\epsilon_k)\right] \epsilon_k v_k.
  \end{equation}
  Here $v_k = d\epsilon_k / dk$  and $f_T(E)$ is the Fermi factor $(e^{E / k_B T} + 1)^{-1}$, and this is of the desired Stefan-Boltzmann form.

At small temperatures compared to the bandwidth of Fermi excitations, so that $x = \epsilon_k / k_B T$ runs from $-\infty$ to $\infty$, then we obtain for a small temperature difference $dT$
\begin{equation}
  {J_E \over dT} = {{k_B}^2 T \over 2 \pi \hbar} \int_{-\infty}^\infty\,dx\, {x^2 e^x \over (e^x + 1)^2} = {\pi^2 {k_B}^2 T \over 3 h},
\end{equation}
which is consistent with (\ref{cft1d}) with central charge $c=1$ as expected.  The universality in~\eqref{cft1d} can be viewed as a generalization of the familiar observation that the ``quantum of thermal conductance'' is the same for fermions and bosons, as one obtains the same value for integral by replacing the Fermi factor by the Bose factor and integrating only from 0 to $\infty$.

\subsection{Non-equilibrium steady states and Fourier's law}

It should be noted that the free case already shows behavior different from the Fourier law expected in a non-integrable system.  In a non-integrable system, left- and right-moving excitations are expected to interact, leading to local thermal equilibrium.  At long times, there will be a smooth change of local temperature from the left reservoir to the right reservoir: Fourier's law predicts a local energy current determined by the temperature distribution $T(x)$ and the linear-response, near-equilibrium thermal conductivity $\kappa$, 
\begin{equation}
j_E(x) = - \kappa(T(x)) \nabla_x T.
\label{fourier}
\end{equation}
In other words, far-from-equilibrium behavior at short times becomes closer to equilibrium behavior as time increases.  The gradient of the temperature distribution reduces as time increases and there is no nonzero long-time value of the local energy current.  Numerical observations by time-dependent DMRG are consistent with this scenario~\cite{karraschilanmoore} although very long times are difficult to access.  Of course, (\ref{fourier}) is exactly what one expects in the hydrodynamical regime of even a classical system, and integrable models can be viewed as exceptions to conventional hydrodynamics -- for energy transport in the XXZ model, this is actually true in a precise sense described below.

There are two more general observations worth making before jumping in to the integrable interacting case, where several new features arise.  
First, we focused in the above on temperature and chemical potential gradients as these are simplest and most experimentally relevant, but from a theoretical point of view this is an oversimplification.  For an integrable system such as the XXZ model, one can ask more generally whether two different generalized Gibbs ensembles (i.e., prepared with an infinite number of Lagrange multipliers, rather than just temperature and chemical potential) prepared in the two reservoirs lead to a steady state that is homogeneous over a central spatial region whose size increases with time.  It seems likely, based on simulations of the XXZ model, that such a steady state does exist, but its details (except in the free case) remain incompletely understood.  Fortunately, certain special properties of energy transport in the XXZ model, now discussed, lead to exact non-perturbative results that can be confirmed numerically, but these do not yet lead to a complete understanding of the non-equilibrium steady state.

Second, as it is not yet known rigorously that a homogeneous steady state exists for long times in the protocol of Fig. 1, it is perhaps unsurprising that it is also unknown how steady states prepared in this manner are related to those prepared by other protocols.  The evolution from reservoir initial conditions is widely used both in the present context and in quantum impurity models~\cite{PhysRevLett.101.140601}, where some effort has gone into justifying its equivalence to experimental situations.  An alternative way of creating steady states is via non-Hermitian boundary conditions on the ends of a finite chain.  Such Lindblad-type boundary conditions for charge (i.e., boundary conditions that add particles to one end and subtract particles to the other end) were important in Prosen's approach to steady states of an open XXZ chain~\cite{prosenxxz}, for example, and have interesting connections to the classical stochastic asymmetric exclusion process.  It is not yet known how to make a non-Hermitian boundary condition equivalent to the thermal reservoirs described here for energy transport, and it would be valuable to understand this non-equilibrium version of the equivalence of ensembles in standard thermodynamics.

\begin{figure}
\begin{center}
\includegraphics[width = 0.75\columnwidth]{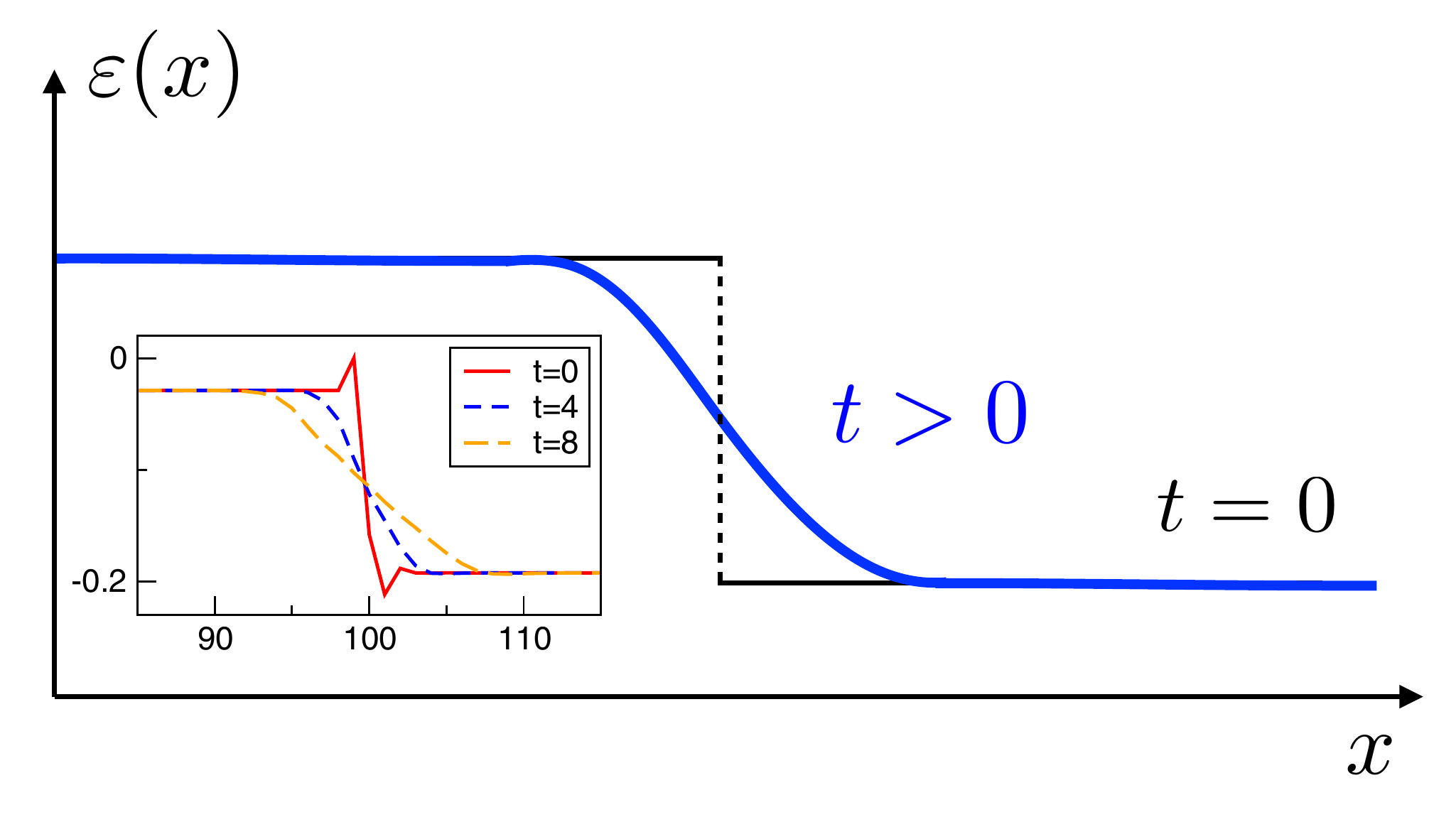}
\vspace{-.2in}
\end{center}
\caption{Spatial spread of energy density $\varepsilon(x)$ after reservoir initial conditions. Inset: example of numerical data from Ref.~\cite{karraschvasseurmoore} obtained using the density matrix renormalization group (DRMG).}
\label{Fig3}
\end{figure}

\subsection{Integrable case and expansion potentials}

In the interacting XXZ model, an important difference emerges between charge and energy transport, which are equivalently straightforward in free theories.  It was observed numerically some years ago~\cite{karraschilanmoore} that a Stefan-Boltzmann law, i.e., a function of one temperature as in (\ref{ffunction}), appeared to exist to numerical accuracy (of order a few percent) in the long-time steady state of energy transport but not for charge transport.  This is surprising, as one would generally expect the thermal current to be a function of $T_L$ and $T_R$ jointly and not just the difference of {\it one} function evaluated at the two temperatures, and indeed this more complicated dependence is what happens in theoretical estimates~\cite{bernarddoyon,PhysRevB.90.161101}.

The reason for this difference is that energy current in the XXZ model has some additional features that make its behavior even far from equilibrium quite universal.  It turns out that a precise theory can be developed for the {\it spatially integrated} energy current over the interaction region, which is quantitatively connected to the first moment of the energy distribution (Fig.~\ref{Fig3}) as might be measured in an experiment with ultracold atoms.  The rate of increase of the integrated energy current in time can be calculated exactly~\cite{karraschvasseurmoore} and verified against the numerical simulations.  We review the basic idea of this calculation here, which uses the fact that energy current in the XXZ model is itself a conserved charge to show that some features normally associated with continuum Lorentz- or Galilean-invariant theories become true even in a lattice model.

In the infinite list of conventional conserved quantities of the XXZ model, there is a special feature: the current of energy is also a conserved charge, {\it i.e.},
\begin{equation}
\left[\int_{-\infty}^\infty j_E\,dx, H\right] = 0,
\label{currentcons}
\end{equation}
and it appears as the density in a continuity equation,
\beq
\partial_t j_E + \partial_x P = 0.
\label{secondcons}
\eeq
This is a stronger statement than simply energy conservation; for example, charge is conserved, so charge density satisfies a continuity equation with charge current, but the charge current is itself not the conserved density in any continuity equation.

\begin{figure}
\begin{center}
\includegraphics[width = 0.75\columnwidth]{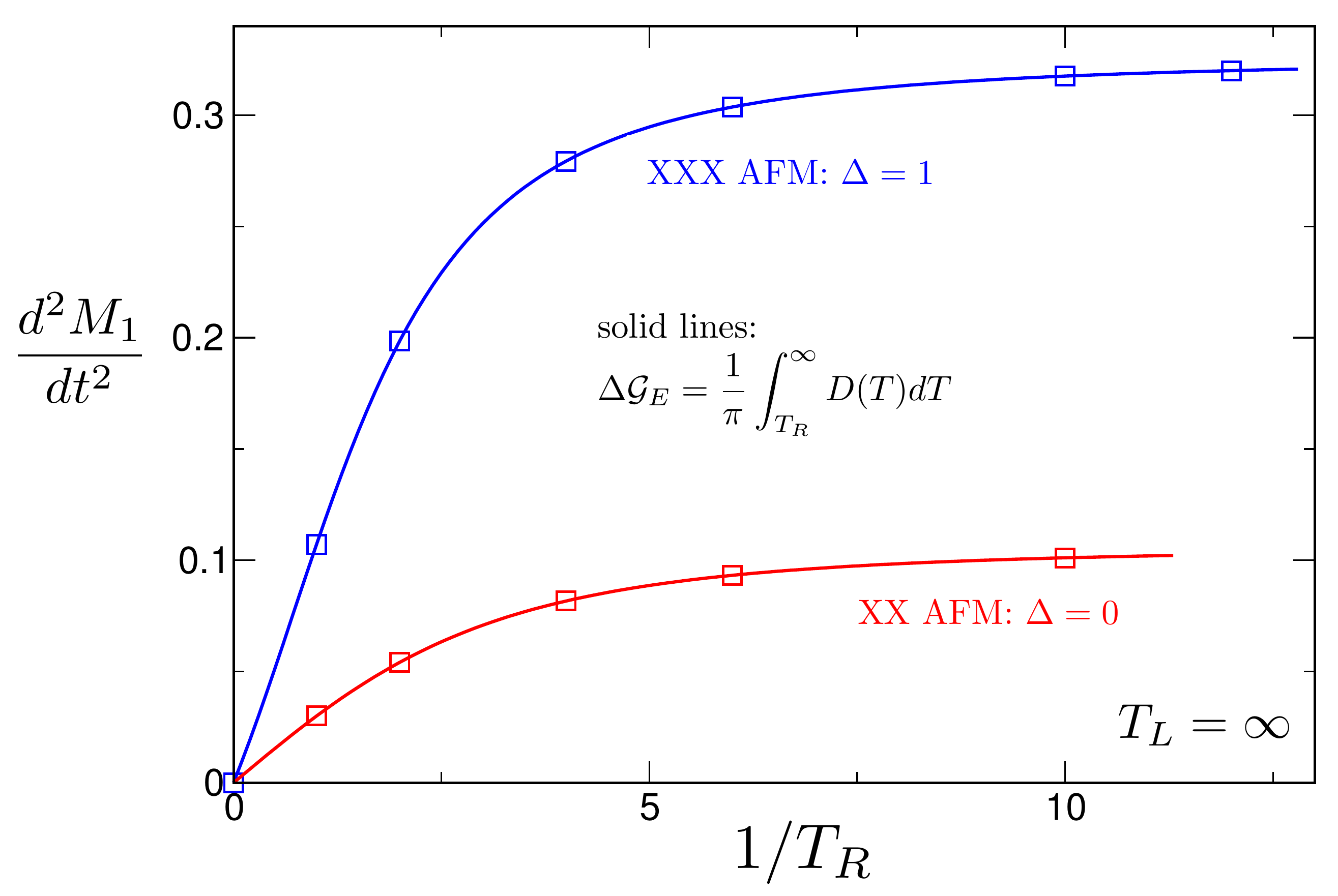}
\vspace{-.2in}
\end{center}
\caption{Increase of the first moment $M_1 = \int  x \varepsilon(x) dx $ of energy density far from equilibrium as a function of $T_R$ with fixed $T_L=\infty$ and $\mu=0$. The DMRG data from Ref.~\cite{karraschvasseurmoore} agrees perfectly with the variation of the expansion potential ${\cal G}_E$ than can be expressed in terms of the thermal Drude weight $D(T)$.  }
\label{FigDMRG}
\end{figure}

There is a rather dramatic consequence of this simple fact: the expansion potential obtained from the thermodynamic average of the ``current of energy current'' ${\cal G}_E(\mu,T) = - \langle P \rangle_{\mu,T}$ governs the increase of integrated energy current, and thereby the increase of the first moment $M_1 = \int x \varepsilon(x) dx  $ of energy density, arbitrarily far from equilibrium~\cite{karraschvasseurmoore}:
\begin{equation}
\frac{d^2 M_1}{d t^2} = \Delta_{R \to L} {\cal G}_E = \int_{R \to L} d {\cal G}_E,
\label{spreading}
\end{equation}
where the integral is over an arbitrary path in the chemical potential and temperature plane.  In other words, there is a Stefan-Boltzmann relation for the spatially integrated current, and the expansion potential plays the role of the Stefan-Boltzmann law (Fig.~\ref{FigDMRG}).

The derivatives of the expansion potential with respect to $\mu$ and $T$ are simply related to the Drude weights for energy current and thermopower, and some relationships between these Drude weights~\cite{PhysRevB.67.224410} that were used to obtain the thermopower exactly~\cite{doi:10.1143/JPSJS.74S.196} are simply the statement that derivatives of ${\cal G}(\mu,T)$ commute; in this sense they are non-equilibrium Maxwell relations.  Predictions like (\ref{spreading}) for the spatially integrated current do not imply without additional assumptions that the point current at the junction, $j_E(x=0)$, also satisfies an equation of the form (\ref{ffunction}), but at least they explain why such a relation might hold approximately for the energy current while being strongly violated for charge currents.

Hence a quick summary of the current state of knowledge for reservoir initial conditions is as follows.  Integrable models tend to have well-defined, homogeneous steady states that occupy a large region of space at long times.  The detailed nature of the steady state is unclear in the interacting case, and there are unanswered basic questions such as whether it is a generalized Gibbs ensemble and, if so, how its parameters are determined by the reservoir parameters.  On the bright side, special features of energy transport in the XXZ model mean that many properties can be understood arbitrarily far from equilibrium in terms of an extra continuity equation and associated expansion potential.

There is also considerable numerical information on currents in the XXZ model, and the validity of the numerics is confirmed even in the interacting case by the high-accuracy agreement for the case of spatially integrated energy currents shown in Fig.~\ref{FigDMRG}.  Two directions for future work are that it should be possible to compare (at least numerically) the steady states created by reservoir and Lindblad-type non-Hermitian boundary conditions, and there are some exciting theoretical predictions for Lorentz-invariant theories~\cite{Bhaseen:2015aa,bernarddoyonunpub} without integrability that might be testable in specific models.  We now turn to the different, more local type of integrability that appears in disordered systems and has equally profound consequences for dynamics.

\section{Quantum dynamics of disordered systems: many-body localization}

\subsection{Dynamics of isolated quantum systems: thermalization {\it vs} localization }

Let us go back to the question of the thermalization of a generic (non-integrable) isolated many-body quantum system after a global quench. As we discussed in the introduction, thermalizing systems have no local memory of their initial condition, so that at long times, they can be efficiently described in terms of a few thermodynamic parameters like temperature. Thermalization should occur independently of the initial state of the system, and it is instructive to consider the case where the system is initialized in a highly excited eigenstate at finite energy density. The dynamics is then trivial, but the thermalization scenario requires that statistical mechanics be encoded in the chosen eigenstate: this is the so-called {\it eigenstate thermalization hypothesis} (ETH)~\cite{PhysRevA.43.2046,PhysRevE.50.888,Rigol:2008kq}. Whereas the system remains in a pure (eigen)state at all times, the eigenstate thermalization hypothesis states that the expectation value of local observables in such an eigenstate is a smooth function of energy, coinciding with the prediction of the microcanonical ensemble at the corresponding energy density. In particular, this implies that the reduced density matrix of a small subsystem should take a Gibbs form, with an effective temperature that depends only on the energy density of the chosen eigenstate. It is worth emphasizing that in thermalizing systems, local observables cannot distinguish between nearby eigenstates in the spectrum, in sharp contrast with the localized systems we will describe below. 
 
  One important consequence of the ETH scenario is that highly excited eigenstates at finite energy density are highly entangled, as the entanglement entropy of a subregion $A$ in eigenstates should coincide with a thermal (thermodynamic) entropy, and should therefore be extensive
\begin{equation}
S_A = s_{\rm th}(\epsilon){\rm Vol}(A),
\end{equation}
where $ s_{\rm th}(\epsilon)$ is the thermodynamic entropy at the energy density of the chosen eigenstate and ${\rm Vol}(A)$ is the volume of the subregion $A$. 
This volume law scaling of entanglement is of course natural for excited states, and contrasts with the area law scaling expected for (gapped) quantum group states. As we saw in the previous section, not all many-body quantum systems thermalize, and integrable systems for example fail to reach thermal equilibrium, leading to interesting new nonequilibrium steady states. However, integrable systems are not generic in that they are highly fine-tuned, and generic perturbations will break integrability. 

Remarkably, in the presence of disorder, it turns out that there exists another generic scenario, {\it many-body localization} (MBL)~\cite{PhysRev.109.1492,FleishmanAnderson,Gornyi,BAA,PalHuse,PhysRevB.75.155111}, for the dynamics of isolated quantum systems, which can be thought of as a complete breakdown of the intuitive picture of thermalization and statistical mechanics~\cite{2014arXiv1404.0686N}. To see this, it is useful to go back to Anderson insulators~\cite{PhysRev.109.1492}, that are characterized by localized (exponentially decaying) single-particle orbitals if disorder is strong enough. In particular, in the following we will be focusing for concreteness on one-dimensional systems of spinless fermions in a random potential
\begin{equation}
H =  \sum_i -t \left(c_{i+1}^\dagger c_i +{\rm h.c.} \right) +  \mu_i n_i,
\label{eqAI}
\end{equation}
where any finite amount of disorder is known to be enough to localize all the single-particle wave functions at {\it all} energies. Such Anderson insulators fail to thermalize because the excitations that would ordinarily move around and transport energy are localized by the disorder, thereby preventing the system from acting as its own heat bath. In particular, the entanglement entropy of the eigenstates of eq.~\eqref{eqAI} satisfies an area law $S \sim {\cal O}(\xi_0)$, with $\xi_0$ the single-particle localization length at the considered energy density, so that ETH is violated. 

A natural objection at this point would be that Anderson insulators such as~\eqref{eqAI} are also integrable, as in fact they are non-interacting systems. However, contrary to integrability, the phenomenon of localization turns out to be generic. Quite remarkably, localization can persist at finite energy density, in highly excited states, even in the presence of moderate interactions if the disorder strength is strong enough. In a seminal work~\cite{BAA}, Basko, Aleiner and Altshuler gave strong arguments for the existence of such many-body localized systems. Further numerical studies~\cite{PalHuse,PhysRevB.75.155111,Luitz}, mostly restricted to the random-field XXZ Hamiltonian
\begin{equation}
H_{\rm XXZ} = \sum_i  J (S^x_i S^x_{i+1}+S^y_i S^y_{i+1}) + J_z S^z_i S^z_{i+1} + h_i S^z_i,
\label{eqXXZrandom}
\end{equation}
with $S_i = \sigma_i^z/2$ (equivalent to interacting spinless fermions after a Jordan-Wigner transformation, with the interaction strength being given by $J_z$), further confirmed the existence of an MBL phase even at {\it infinite temperature} ({\it i.e.} extending throughout the full many-body spectrum) at strong enough disorder. One of the key features of MBL systems is that excited-eigenstates satisfy a area-law scaling of the entanglement entropy~\cite{BauerNayak,PhysRevLett.111.127201} ($S\sim {\rm const}$ in 1D), so that they clearly violate ETH. Intuitively, such MBL eigenstates ``look like'' gapped groundstates, and can have properties akin to zero-temperature quantum groundstates, but now at finite energy density. This will turn out to have especially interesting consequences that we will discuss in the following, with a focus on universal dynamical features. 

Before we go further, let us emphasize that the following discussion is not meant as a comprehensive review of all the recent developments the field of many-body localization, but rather as a partial (and subjective) account of some of the exciting properties of MBL systems from the perspective of universal quantum dynamics. We refer the interested reader to the excellent reviews~\cite{2014arXiv1404.0686N,doi:10.1146/annurev-conmatphys-031214-014701} for more details. Among the topics we will not cover here, let us mention the recent experimental realizations of many-body localized systems~\cite{Schreiber842,2015arXiv150807026S}, MBL driven (Floquet) systems~\cite{PhysRevLett.115.030402,PhysRevLett.114.140401,Ponte2015196,2015arXiv150803344K}, MBL systems coupled to a bath~\cite{PhysRevB.90.064203,MeanFieldMBLTransition,PhysRevB.92.245141,2016arXiv160107184H}, MBL with long-range interactions~\cite{PhysRevLett.113.243002,PhysRevB.91.094202, PhysRevB.92.134204}, or (quasi-)MBL in translation-invariant systems~\cite{2013arXiv1307.2288G,2013arXiv1309.1082S,2014arXiv1405.3279D,2014arXiv1405.5780H,2014arXiv1410.7407Y}.

\subsection{Universal dynamics of many-body localized systems}

The nonequilibrium dynamics of (noninteracting) Anderson insulators after a global quench is relatively simple. Let us imagine starting from a random $S_z$ product state (or, say, a N\'eel state), and look at the time evolution of the bipartite entanglement and of the variance of the total spin on half of the chain, characterizing transport. Both quantities behave similarly after such a global quench, as they both grow and quickly saturate, corresponding to the expansion of wave packets to a size of the order of the localization length, after which particles ``get stuck'', so that quantum information and conserved quantities stop spreading. 
Many-body localized systems, on the other hand, have a much richer universal dynamics. Whereas there is no energy or charge (spin) transport, the bipartite entanglement entropy was observed numerically to grow logarithmically~\cite{PhysRevB.77.064426,PhysRevLett.109.017202,PhysRevLett.110.260601}, reaching a finite value proportional to the size of the system, corresponding to volume-law (but non thermal) entanglement. This unbounded growth of entanglement for an infinite system initially appeared hard to reconcile with the absence of transport and of thermalization. This key observation led to what is now known as the ``local integrability'' picture of the MBL phase~\cite{PhysRevLett.110.260601,PhysRevLett.111.127201,PhysRevB.90.174202,VoskAltmanPRL13,2013arXiv1307.0507S}. The idea is that fully many-body localized systems -- MBL systems where there is no mobility-edge and where (almost) all the eigenstates are MBL and satisfy an area-law entanglement scaling -- have an emergent, generic ``integrability'' in the sense that they possess infinitely many (quasi)local integrals of motions. More precisely, starting for concreteness from eq.~\eqref{eqXXZrandom} at strong disorder, it is possible to define a set of localized spins $\tau^z_i$, or ``l-bits'', that can be thought of a a dressed version of the physical spins, such that $[\tau^z_i,H]=0$ and $[\tau^z_i,\tau^z_j]=0$. These l-bits are thus conserved quantities, and they are local in the sense of being expressed in terms of order $\sim {\cal O}(\xi)$ physical spins (with exponential tails), with $\xi$ the many-body localization length. Note these conserved quantities are (quasi)local in a sense that is very different from integrable systems, where the conserved quantities are sum of local (or quasi-local~\cite{prosenxxz,PhysRevLett.111.057203,2016arXiv160300440I}) operators. In the non-interacting case, $\tau^z_i$ is simply the occupation number of the single particle orbital localized around the site $i$, and in the interacting case the l-bits can be thought of as a generalization of the concept of single particle orbitals, ``dressed'' by interactions. We emphasize that this infinite set of local integrals of motion is {\it generic}, in sharp contrast with traditional integrable systems. 

The Hamiltonian~\eqref{eqXXZrandom} expressed in terms of l-bits then takes the simple form~\cite{PhysRevLett.111.127201,PhysRevB.90.174202}
\begin{equation}
H = \sum_i  \epsilon_i \tau_i^z + \sum_{ij}  J_{ij} \tau_i^z \tau_j^z +  \sum_{ijk}  K_{ijk} \tau_i^z \tau_j^z \tau_k^z + \dots,
\label{eqlbits}
\end{equation}
where $\tau^{\pm}_i$ terms are not allowed since $[\tau^z_i,H]=0$, and $\tau_i^z=U^\dagger \sigma_i^z U$ with $U$ a quasi-local unitary transformation (see~\cite{Ros2015420} for an explicit construction). Note that the first term in the expansion corresponds to a non-interacting Anderson insulators whereas higher-order terms are due to interactions. 
This Hamiltonian~\eqref{eqlbits}  can be considered as a phenomenological model for MBL phase~\cite{PhysRevLett.111.127201,PhysRevB.90.174202} (later made rigorous by Imbrie~\cite{2014arXiv1403.7837I} with minimal assumptions), or alternatively, following Vosk and Altman~\cite{VoskAltmanPRL13}, as a dynamical renormalization group (RG) fixed point describing the MBL phase. This ``fixed-point'' Hamiltonian is then reminiscent of a real space version of a Fermi liquid, where the typical value of the interaction coefficients decay exponentially with distance, {\it e.g.} $J_{ij} \sim J_z {\rm e}^{-|i-j|/{\xi_0}}$ (recall that $J_z$ represents the strength of the interactions), reflecting the localized nature of the phase.   

This local integrability picture explains most of the properties of MBL systems. For instance, the lack of thermalization follows immediately from the very existence of the conserved quantities $\tau_i^z$, while the area-law scaling of the entanglement in eigenstates can be understood as the ``dressing'' of the l-bits in terms of the physical spins --- the eigenstates in the l-bit basis being simply $\tau^z$ product states. Whereas the eigenstates of eq.~\eqref{eqlbits} in the l-bit basis are trivial, the dynamics is slightly more subtle. Let us focus on only two l-bits $i$ and $j$ with Hamiltonian $H=  \epsilon_i \tau_i^z +  \epsilon_j \tau_j^z + J_{ij} \tau_i^z \tau_j^z$ where we dropped the higher-order terms for the sake of simplicity. Starting from a generic ({\it i.e.} not an eigenstate) initial product state, it is straightforward to observe that the entanglement entropy $S(t)$ of the spin $i$  increases with time because of the interaction term   $J_{ij} \tau_i^z \tau_j^z$ in $H$ that will induce some dephasing between the two spins, which will be maximal around $t J_{ij} \sim 1$~\cite{PhysRevLett.110.260601}. This can be easily generalized to argue that l-bits separated by a distance $r$ will eventually dephase on time scales of order $t J_z {\rm e}^{-r/\xi_0} \sim 1$, contributing to the growth of the entanglement entropy~\cite{PhysRevLett.110.260601}
\begin{equation}
S(t) \sim \xi_0 \log \left( J_z t \right),
\label{eqLogGrowth}
\end{equation}
in perfect agreement with numerical observations. 
 
 \subsection{Experimental probes of the dephasing nature of many-body localized systems}
 
 Combined with the absence of transport and thermalization, this slow, logarithmic growth of entanglement in MBL systems after a global quench is often considered as smoking gun of the ``dephasing but non dissipative'' nature of many-body localization described in the previous section, allowing one to distinguishing MBL from Anderson insulators. It is therefore crucial to find realistic ways to measure this entanglement growth experimentally, with all the usual difficulties associated with measuring entanglement entropy. Thankfully, the ``dephasing'' physical mechanism underlying the slow entanglement growth~\eqref{eqLogGrowth} has in fact  little to do with entanglement, and shows up in various observables that are, in principle, much simpler to access experimentally~\cite{PhysRevLett.113.147204,PhysRevB.91.140202,PhysRevB.90.174302}. 
 
  \begin{figure}[t!]
\begin{center}
\includegraphics[width = 0.95\columnwidth]{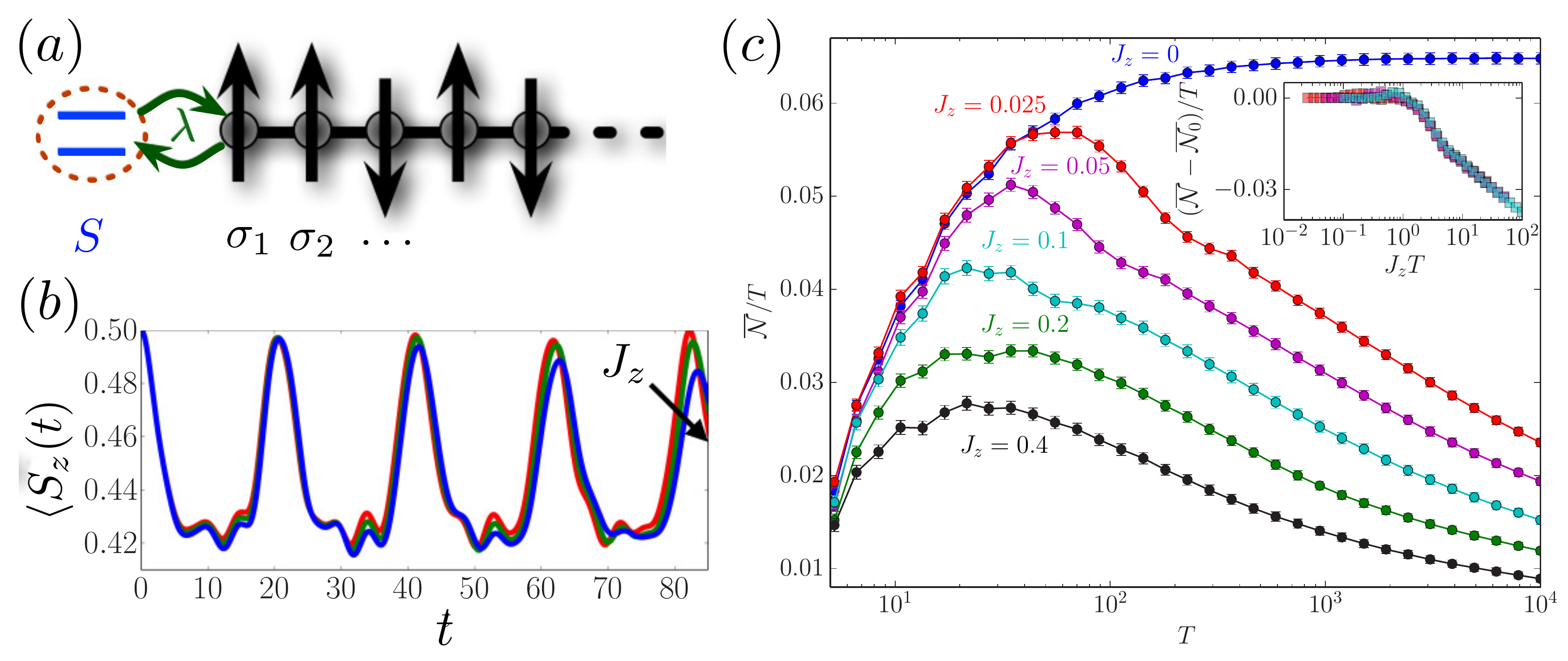}
\vspace{-.2in}
\end{center}
\caption{{\bf Revivals of local observables in a many-body localized system.} (Figures from Ref.~\cite{PhysRevB.91.140202}) (a) Setup: we consider the post-quench dynamics of a two-level `qubit' $\mathbf{S}$ coupled to a one-dimensional chain of atoms.  (b) Time series for a single instance of disorder, and influence of the interactions strength $J_z$ on the revivals. (c) Disorder-averaged revival rate $\overline{\mathcal{N}(T)}/T$ as function of total time, $T$.  Upon adding interactions of strength $J_z $, revivals are suppressed beyond $T^* \sim J_z ^{-1}$. (Inset) The same data collapses onto a universal curve when plotted against $J_z  T$, with ${\cal N}_0(T) = \left. {\cal N} (T) \right|_{J_z=0}$.}
\label{fig:Revivals}
\end{figure}
 
 For example, it was argued that MBL could be characterized by the revival (defined as the return of a time-dependent observable sufficiently close to its initial value after deviating sufficiently far from it) rate of local observables~\cite{PhysRevB.91.140202}, or more specifically of a single ``probe'' qubit locally coupled to a one-dimensional reservoir being either ergodic, Anderson localized or MBL. For an ergodic reservoir that would act as a bath for the qubit, an approximate revival of the wave function requires synchronizing $\sim 2^L$ frequencies with $L$ the size of the system, so that revivals quickly become extremely unlikely. For a Anderson insulator $H=\sum_i \epsilon_i \tau_i^z$ however, the qubit effectively sees a system of size of the order of the localization length $\xi$, and since the (free fermion) spectrum is additive, the revivals of the qubit are governed only by the frequencies  $\epsilon_i$ of the  ${\cal O}(\xi)$ l-bits $\tau_i^z$ closest to the probe qubit. The revivals are therefore controlled by $N_0 \sim \xi$ frequencies in that case so that we expect the asymptotic revival rate (number of revivals per time unit) to be constant and to scale as $\Gamma_{0} \sim {\rm e}^{-N_0} \sim {\rm e}^{-\xi}$, ignoring unimportant prefactors. Finally, if the reservoir is many-body localized with interactions $J_z$, we expect interactions to start inducing dephasing after a time scale $t \sim J_z^{-1}$, thereby slowly destroying the revivals of the non-interacting case. However, the probe qubit does not immediately ``see'' $2^L$ or $2^\xi$ frequencies, but instead, very much like the argument leading to eq.~\eqref{eqLogGrowth}, the effective number of frequencies controlling the approximate revivals of the probe qubit can be argued to grow logarithmically $N(t) \approx N_0 + \alpha \log  t$ at strong enough disorder. This implies that the constant qubit revival rate in the Anderson insulator is changed to a universal logarithmic decay (resulting from expanding the power-law scaling of $\Gamma \sim {\rm e}^{-N(t)}$ at strong enough disorder) upon adding interactions. This logarithmic decay can be precisely related to the dephasing mechanism responsible for the slow, logarithmic growth of entanglement in the MBL phase and thus provides a quantitative, experimentally observable alternative to entanglement growth as a measure of the ``nonergodic but dephasing'' nature of many-body localized systems (Fig.~\ref{fig:Revivals}). Similar probes have been proposed in the literature, relying on spin echo techniques~\cite{PhysRevLett.113.147204} or relaxation after a quantum quench~\cite{PhysRevB.90.174302} (see also~\cite{PhysRevB.92.180205,2016arXiv160102666G} for more recent proposals).

\subsection{Excited-state quantum criticality}

As we already mentioned above, a key property of a (fully) many-body localized system is the area-law scaling of entanglement satisfied by (almost all) the eigenstates. In other worlds, the excited eigenstates of an MBL system ``look like'' gapped quantum groundstates. This remarkable property provides us with some intuition as to why MBL systems can host completely new phenomena at finite energy density, that would ordinarily occur only at zero temperature, in quantum groundstates. MBL indeed leads to the counterintuitive property that disorder can actually protect symmetry-breaking~\cite{HuseMBLQuantumOrder,PhysRevLett.113.107204,PekkerRSRGX}, topological ~\cite{HuseMBLQuantumOrder,BauerNayak} or symmetry protected topological~\cite{PhysRevB.89.144201,BahriMBLSPT,CenkeMBLSPT,ACPMBLSPT} orders at finite energy density, even in regimes where such orders are forbidden by statistical mechanics. In particular, the usual Mermin-Wagner or Peierls type arguments that would imply that thermal fluctuations would destroy quantum order simply do not apply here since they rely on statistical mechanics. For example, the domain walls associated with symmetry-breaking quantum order in one-dimension that would proliferate at finite temperature and destroy quantum order can be localized by disorder, leading to the existence of distinct 1D MBL phases at finite energy, with or without spontaneous discrete symmetry breaking in eigenstates~\cite{HuseMBLQuantumOrder}. A concrete example is provided by the transverse field Ising chain~\cite{PhysRevLett.113.107204,PekkerRSRGX} 
\begin{equation}
H = -\sum_i J_i \sigma^z_i \sigma^z_{i+1} + h_i  \sigma_i^x + J_i^\prime \sigma^x_i \sigma^x_{i+1}, 
\label{eqIsing}
\end{equation}
where we have included weak interactions ($J_i^\prime \neq 0$) preserving the ${\mathbb Z}_2$ symmetry in order to make the system non-integrable. For strong enough disorder, there are two distinct MBL phases. In the regime where $J_i$ dominates ($\overline{\log J_i} \gg \overline{\log h_i}$ where $\overline{O}$ refers to averaging $O$ over disorder), the local conserved quantities are dressed Ising terms $\tau^z_i = U^\dagger \sigma_i^z \sigma^z_{i+1}U$, and the eigenstates come in almost degenerate Schr\"odinger cat pairs  $\Ket{n}_\pm = (\Ket{n} \pm {\mathcal C}  \Ket{n})/\sqrt{2}$ that are even/odd under the ${\mathbb Z}_2$ symmetry generated by ${\mathcal C} = \prod_i \sigma_i^x$, where $\Ket{n}$ is some eigenstate-dependent pattern of $\sigma^z$ magnetization breaking the ${\mathbb Z}_2$ symmetry. The energy splitting between the two true eigenstates $\Ket{n}_\pm$ is exponentially small in system size and scales as $\sim {\rm e}^{-L/\xi}$ with $\xi$ the localization length, implying that the broken-symmetry state $\Ket{n}$ becomes metastable in the limit of large systems. This is very similar the usual scenario of spontaneous symmetry breaking in the groundstate, but now occurring in highly excited eigenstates. Note that the symmetry is broken is a spin-glass manner, since the correlator $\langle \sigma_i^{z} \sigma_{i+r}^{z} \rangle $ evaluated in an eigenstate changes sign every time $r$ crosses one of the (localized) domain walls -- unlike a ferromagnet where all spins would all point in the same direction. This spin-glass MBL phase can be diagnosed by a non-vanishing value of the Edwards-Anderson order parameter $m_{\rm EA} = 1/L^2 \sum_{i \neq j} \langle \sigma_i^{z} \sigma_{j}^{z} \rangle^2$ in the thermodynamic limit, averaged over eigenstates and over disorder configurations. Up to a Jordan-Wigner transformation, this corresponds to the survival of Majorana edge zero modes in excited states, at finite energy density in the presence of interactions. In the opposite regime where the transverse field $h_i$ dominates ($\overline{\log h_i} \gg \overline{\log J_i}$),  the local conserved quantities are dressed transverse field terms $\tau^z_i = U^\dagger \sigma_i^x U$, and the symmetry remains unbroken with short-range $\sigma^z$ correlations, so that the system is in a many-body localized paramagnetic phase. 

 \begin{figure}[t!]
\begin{center}
\includegraphics[width = 0.95\columnwidth]{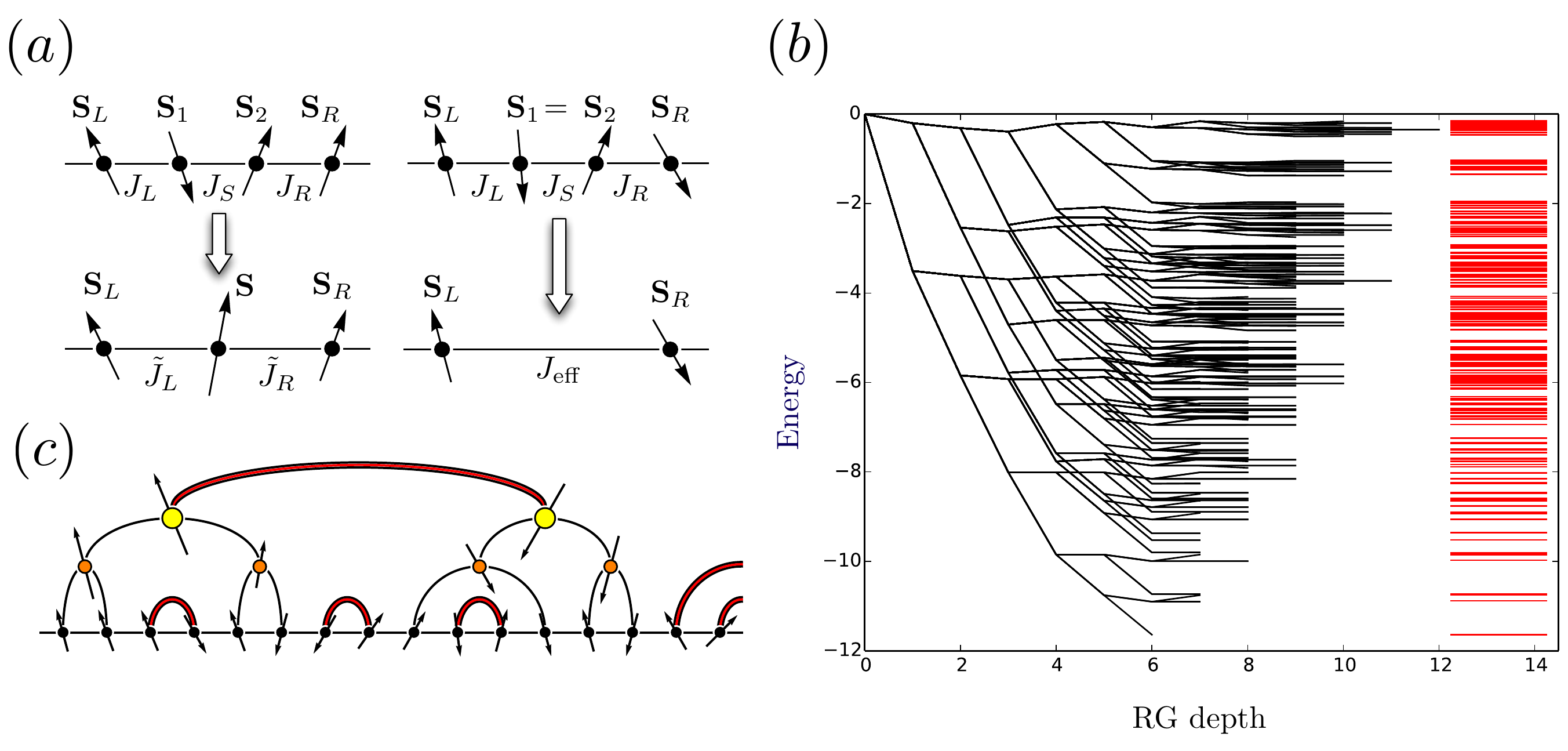}
\vspace{-.2in}
\end{center}
\caption{{\bf Quantum critical glasses and critical points between different MBL phases.} (Figures from Ref.~\cite{QCGPRL}) (a) General structure of the renormalization process and of the decimation rules. (b) Example of renormalization group ``spectral tree'' where each branch corresponds to an eigenstate with a different energy, constructed for a disordered Fibonacci chain~\cite{PhysRevLett.98.160409} and compared against exact diagonalization results in red. (c) Random singlet structure of the eigenstates, where ``spins'' (anyons) of various sizes are paired into singlets. }
\label{fig:PhaseRSRG}
\end{figure}

The universal properties of such excited-state critical points (and critical phases) separating area-law entangled MBL phases can be efficiently captured by strong disorder real space renormalization group (RSRG) approaches~\cite{PekkerRSRGX,VoskAltmanPRL13,PhysRevLett.112.217204,QCGPRL}. These new approaches essentially rely on the RSRG procedure that has proven very useful to study zero-temperature antiferromagnetic random spin chains~\cite{PhysRevLett.43.1434,PhysRevB.22.1305,BhattLee,FisherRSRG1,FisherRSRG2,WesterbergPRL}, in which strong couplings in the Hamiltonian are decimated before weaker ones, putting the spins involved in the strongest coupling in their local groundstate. In the example~\cite{FisherRSRG1} of the Ising chain~\eqref{eqIsing} (where we consider the noninteracting case $J^\prime=0$ for the sake of simplicity), if the strongest coupling is a transverse field $h_i=\Omega$, we will first ignore the rest of the chain (since this largest coupling will be typically much larger than its neighbors) and put the spin $i$ in the groundstate $\Ket{\rightarrow}_i=\frac{1}{\sqrt{2}}(\Ket{\uparrow}_i+\Ket{\downarrow})_i$ of the strong transverse field, and then deal with the rest of the chain perturbatively. In our example of a strong transverse field, quantum fluctuations induce an effective Ising coupling $\tilde{J} = J_{i-1}J_{i}/\Omega$ between the neighboring spins $i-1$ and $i+1$. The case of a strong Ising coupling $J_i$ can be dealt in a similar way and the Hamiltonian conserves its form upon renormalization, and weak interactions ($J^\prime_i \neq 0$) can be argued to be irrelevant and do not change the resulting picture. The effective disorder strength grows upon renormalization so that the resulting RSRG flows to infinite randomness~\cite{FisherRSRG1,FisherRSRG2} and is said to be asymptotically exact -- meaning that it is believed to yield exact results for universal quantities such as critical exponents. This method was applied successfully to study the low energy properties of a variety of antiferromagnetic spin chains, including the Heisenberg chain where the resulting groundstate is made of singlets of various sizes. It turns out that many infinite-randomness critical points in 1D can be understood as a property of ``random singlet'' groundstates, including the Ising example above where the groundstate is made of ``singlets'' of Majorana fermions~\cite{Bonesteel}.

This approach was recently generalized to target many-body excited states by observing that at each step, it is possible to project the strong bond onto an excited-state manifold~\cite{PekkerRSRGX}. Coming back to our example of the Ising model, if the strongest coupling at a giving step is a transverse field $h_i$, we may choose to project the corresponding spin onto the excited state $\Ket{\leftarrow}_i=\frac{1}{\sqrt{2}}(\Ket{\uparrow}_i-\Ket{\downarrow}_i)$, so as to increase the total energy of the state. Quantum fluctuations then induce effective renormalized couplings as in the groundstate case, the only difference being that each time a strong coupling is decimated, we can either choose to minimize or maximize its corresponding energy.  The resulting excited-state RSRG (RSRG-X) iteratively resolves smaller and smaller energy gaps, corresponding to slow modes in the dynamics~\cite{VoskAltmanPRL13,PhysRevLett.112.217204}, and allows one to construct, in principle, all the many-body eigenstates of the system~\cite{PekkerRSRGX,QCGPRL} (see also~\cite{2015arXiv150906258M,2015arXiv150803635Y,PhysRevB.93.134207}). This method was applied to an infinite family of 1D spin chains~\cite{QCGPRL} (or more precisely disordered ``anyon" chains~\cite{PhysRevLett.98.160409,Trebst01062008}) that can be intuitively thought of as deformations of the Heisenberg $SU(2)$ chain where the number of irreducible representations (the largest value of the spin) has been truncated. Such anyon chains provide convenient lattice regularizations of the minimal models of conformal field theories for uniform couplings~\cite{Trebst01062008}, and their disordered versions correspond~\cite{FidkowskiPRB09} to the so-called Damle-Huse infinite randomness fixed points~\cite{DamleHuse} at zero temperature. Quite remarkably, the highly-excited eigenstates of such 1D models also have quantum critical properties that are in general different from their equilibrium, zero temperature counterparts.
 
The general structure of the renormalization steps is quite simple (Fig.~\ref{fig:PhaseRSRG}a): as a strong bond between the spins ${\bf S}_1$ and ${\bf S}_1$ is decimated, the energy levels of the corresponding local Hamiltonian can be labelled by the fusion channels of ${\bf S}_1 \otimes {\bf S}_2$. For example, if  ${\bf S}_1={\bf S}_2={\bf \frac{1}{2}}$, there are two fusion channels ${\bf 0} \oplus {\bf 1}$ with different energies, a singlet and a triplet. If the strong bond spins fuse to a singlet, then they drop out of future stages of the RG and virtual excitations of this frozen singlet then mediate effective coupling between the neighboring spins. Another possibility is that the strong bond spins fuse to a new effective ``superspin'' (like the spin one triplet in the example ${\bf \frac{1}{2}} \otimes {\bf \frac{1}{2}} ={\bf 0} \oplus {\bf 1}$)  which continues to participate in later stages of the RG and interacts with its neighboring spins via renormalized couplings. It is useful to think of the outcome of RSRG-X as a tree of possibilities (Fig.~\ref{fig:PhaseRSRG}b), showing the energy of a state as a function of the number of RG steps (number of couplings decimated), where each node corresponds to a choice at a given step (either minimizing or maximizing the energy of the strongest coupling in the example of the Ising model), and each branch corresponds to a many-body eigenstate. The resulting RG tree can then be sampled numerically~\cite{PekkerRSRGX} by a Monte Carlo procedure to obtain physical properties, or analytically~\cite{QCGPRL}  by writing down flow equations for the properties of a typical eigenstate in the middle of the many-body spectrum, corresponding effectively to infinite temperature. The technical details of this RSRG-X procedure are beyond the scope of this review, and we briefly discuss instead the physical properties of the resulting critical points (and more generally, critical phases), dubbed ``quantum critical glasses'' (QCG) in Ref.~\cite{QCGPRL} .

RSRG-X predicts that the eigenstates of these random anyon chains (which include the Ising chain~\eqref{eqIsing} for example)  have a random singlet structure where the effective spins created upon renormalization can grow (like the effective spin one obtain by fusing two spins $1/2$ for example), but are eventually paired in singlets of various ranges (Fig.~\ref{fig:PhaseRSRG}c). This implies that many of the quantum critical properties ordinarily akin to zero-temperature infinite-randomness groundstates now appear in highly excited states, near the middle of the spectrum. In particular, such QCG eigenstates are characterized by a non-thermal logarithmic scaling of the entanglement entropy $S \sim \log L$ (see also~\cite{YichenJoel}), as in infinite randomness groundstates in 1D~\cite{RefaelMoore,Bonesteel,FidkowskiPRB08}. This shows that QCG are non-ergodic since they violate ETH which would imply $S \sim L$ in 1D, while being fundamentally different from MBL which would have area-law entanglement $S \sim {\rm const}$.
They also show power-law average correlations and a glassy scaling of the entanglement growth after a global quench, $S(t) \sim (\log t)^\alpha$~\cite{VoskAltmanPRL13,PhysRevLett.112.217204}, where $\alpha>1$ (recall that $\alpha=1$ for an MBL system, see also Ref.~\cite{2015arXiv151203388L} for a detailed review of the entanglement growth in various disordered systems). The universal properties of these quantum critical glass fixed points are generically distinct from their ground state equilibrium counterparts, and represent novel nonequilibrium critical phases of matter. QCG are a clear example of universality at very high energy density, far away from equilibrium, providing us with a quantum critical analog of MBL. We note that phase transitions between MBL phases remain very poorly understood in general beyond the examples mentioned above, and it would be very interesting to investigate their general nature and their physical properties in higher-dimensional systems. 

The stability of such QCG ``phases'' remains an important question: whereas RSRG-X flows to infinite randomness similarly to the analog $T=0$ RSRG, it essentially assumes the existence of a localized, non-ergodic phase by ignoring resonances that could lead to thermalization. When a strong coupling is diagonalized (``decimated'') thereby resolving a large gap in the spectrum, the corresponding spin(s) is (are) frozen in a given state and  RSRG-X assumes that there is no back action on the frozen spin(s) from weaker couplings which are decimated later on in the procedure. Whereas a formal proof of stability as in the MBL case~\cite{2014arXiv1403.7837I} is unavailable at this point, some arguments suggest that resonant processes ignored by RSRG-X become irrelevant near the infinite randomness fixed point~\cite{VoskAltmanPRL13, PVPtransition}, and a pragmatic approach is to consider RSRG-X as a useful tool to analyze the strong disorder phase assuming localization, which at the very least would control the crossover towards thermalization if resonances turn out to be relevant.   

 \begin{figure}[t!]
\begin{center}
\includegraphics[width = 0.65\columnwidth]{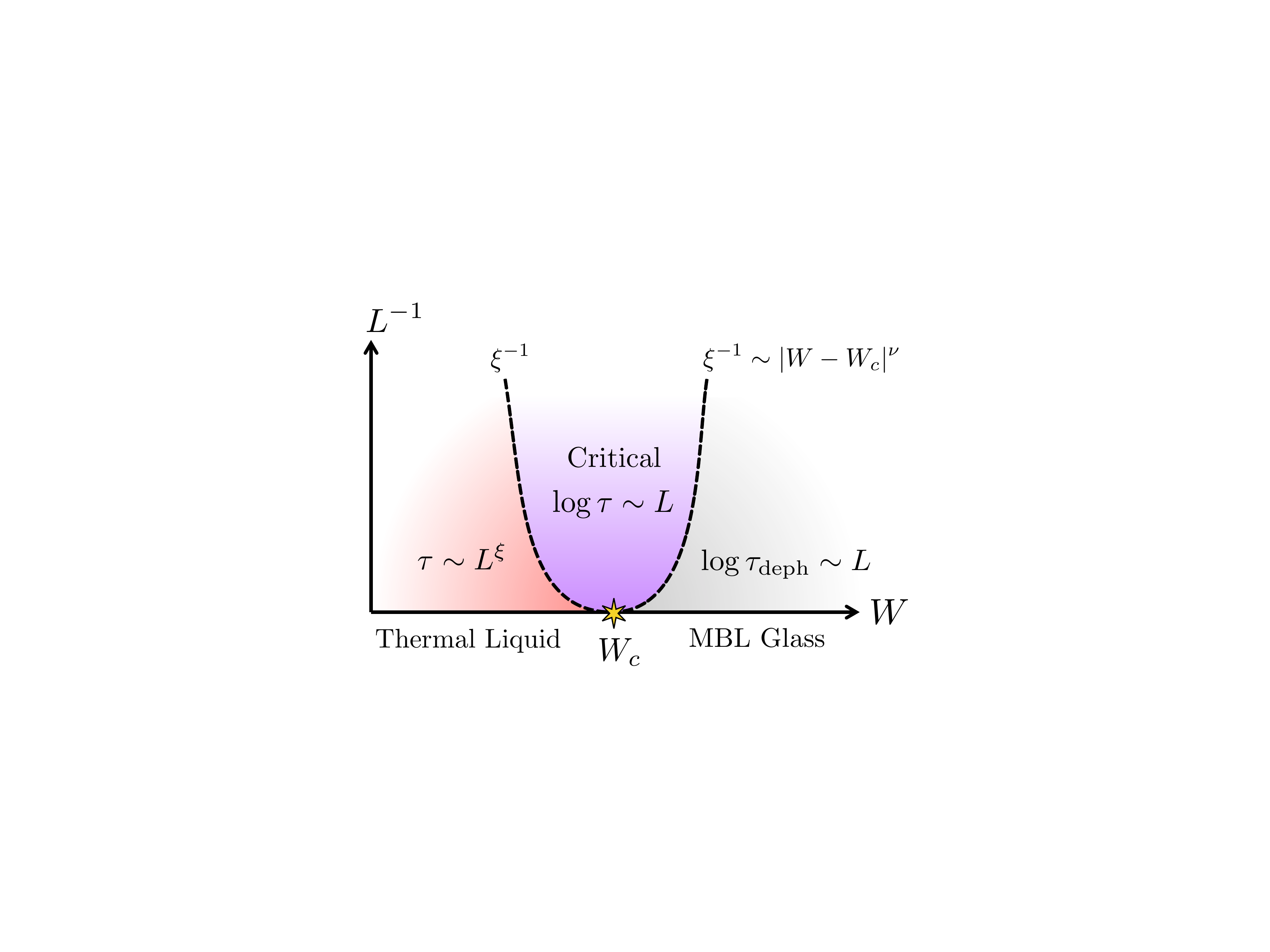}
\vspace{-.2in}
\end{center}
\caption{{\bf Phase diagram and finite size crossovers of the MBL-ETH transition.} (Figure from Ref.~\cite{PVPtransition}) In the thermodynamic limit, there is a continuous phase transition at disorder strength $W=W_c$ characterized by a diverging correlation length $\xi~\sim |W-W_c|^{-\nu}$. In the MBL phase, there is no transport but there is a slow dephasing $L\sim \log \tau_{\rm deph} $ (where the subscript ``deph'' emphasizes that this is the dephasing time rather than the energy transport time scale, which is infinite in the MBL phase) responsible for the logarithmic growth of entanglement. Energy spreads subdiffusively in the ergodic phase with a continuously evolving dynamical exponent $z \sim \xi$, whereas both entanglement and energy spread as logarithmically ($L \sim \log \tau$) at the critical point.     }
\label{fig:PhaseDiagramMBLETH}
\end{figure}

\subsection{Many-body (de)localization transition}

In the previous section, we focused on the excited-state (dynamical) phase transitions (and critical phases) that can occur at strong disorder, separating distinct MBL phases. An even more unusual example of excited-state phase transition is provided by the so-called MBL transition that occurs as a function of disorder strength, for fixed interactions, separating the MBL regime at strong disorder from an ergodic, delocalized thermal liquid phase at weak disorder. We will focus on the ``infinite temperature'' limit by considering eigenstates in the middle of the many-body spectrum. At the MBL transition, the entanglement structure of many-body eigenstates changes dramatically, going from an area-law, quantum groundstate-like entanglement scaling on the MBL side to a volume law scaling in the ergodic (ETH) phase. This phase transition is very unique in that respect, and is fundamentally different from ordinary quantum critical and classical (thermal) phase transitions that separate states with the same entanglement structure -- with area-law and volume-law scaling, respectively. It is also instructive to think of this transition as a fundamental frontier between quantum mechanics, represented by the MBL phase, and the classical equilibrium world, corresponding to the ergodic ETH phase that will effectively thermalize to very high temperature and behave effectively classically.   

Investigating such MBL transitions theoretically requires facing a daunting combination of non-equilibrium dynamics, disorder and interactions, and remains a major challenge. The transition now occurs at finite disorder so that the infinite-randomness RG schemes described in the previous section breakdown, and even the notion of renormalization group for such transitions has to be reconsidered, since they occur in a regime of very high energy, where ordinary thermalizing systems would show no sign of criticality or universality. Moreover, numerical methods that were remarkably successful at helping forging some physical intuition of the MBL phase itself most likely fail near the critical point, since the transition appears to be driven by long-distance properties whose characteristic length scale (correlation length) diverges as the critical point is approached. For example, recent numerical studies~\cite{PhysRevLett.113.107204,Luitz} obtained critical exponents that contradict fundamental bounds~\cite{Chayes,2015arXiv150904285C}, which might suggest that the systems studied numerically are too small (of order $\sim 20$ spins) to access the true scaling universal regime of the transition. Recently, two phenomenogical renormalization group approaches~\cite{VHA,PVPtransition} were proposed (see also~\cite{2016arXiv160302296Z}) to describe the MBL transition in one-dimensional systems, both based on very different physical ingredients. The first approach~\cite{VHA} relies on the scaling of a parameter generalizing the concept of Thouless conductance to the many-body case (see also~\cite{PhysRevX.5.041047}) in a coarse-grained model made of ergodic and localized blocks. The second approach~\cite{PVPtransition} adopts a more microscopic perspective and iteratively (and approximately) constructs the collective many-body resonances that destabilize the MBL phase as disorder is weakened, assuming a self-similar hierarchical structure of these resonances near the critical point. We will not go into the technical details of these two RG approaches here, but we will briefly discuss some aspects of the rich scaling structure of this transition that has emerged in the past two years. 

The transition appears to be continuous (second order), with a diverging correlation length $\xi~\sim |W-W_c|^{-\nu}$ as a function of disorder $W$, with $W_c$ the critical disorder strength separating the ergodic (ETH) regime ($W<W_c$) from the MBL phase ($W>W_c$). The exponent $\nu$ was found to be $\nu \sim 3 - 3.5$, very different from numerical results~\cite{Luitz} (that found an exponent $\nu\sim 0.9$) but in agreement with the fundamental bound $\nu \geq 2$~\cite{Chayes,2015arXiv150904285C}. The dynamics at the critical point $W=W_c$ was predicted (and soon after confirmed numerically~\cite{PhysRevX.5.041047}) to be ``glassy'', with both entanglement and conserved quantities (like energy) spreading as $\sim \log t$. This contrasts with the MBL phase where entanglement also grows logarithmically, but where conserved quantities are localized. Despite being effectively classical and at very high temperature, the thermal (ergodic) phase at weak disorder also shows some signatures of the proximity of the MBL transition. It exhibits a broad regime of anomalously slow subdiffusive equilibration dynamics and energy transport, where entanglement spreads sub-ballistically, and the dynamics of conserved quantities are subdiffusive~\cite{VHA,PhysRevLett.114.100601,Agarwal,PVPtransition,PhysRevB.93.060201}. In particular, energy transport in the ergodic phase is characterized by the the following scaling between time and distances
\begin{equation}
L \sim t^{1/z(W)}, \ {\rm with} \ z(W) \sim \left| W - W_c \right|^{-\eta},
\end{equation}
where the ``dynamical exponent'' $z(W)$ ($z=2$ for diffusion) is continuously evolving and nonuniversal, and diverges at the transition $z = \infty$. This relation should match the scaling $L \sim \log t$ at the critical point on the scale of the correlation length $L \sim \xi $, so that up to logarithmic corrections, $z \sim \xi$ and $\eta = \nu$. This is an example of scaling relation between (universal) critical exponents, very far away from equilibrium and at very high energy. This subdiffusive behavior can be understood easily as a Griffiths effect due to rare events: the key idea is that even in the thermal phase, there are regions that can be locally more disordered and that will behave as localized (or critical) regions. While such insulating regions much larger than the correlation length are exponentially rare, their contribution to transport is exponentially large: they act as bottlenecks for transport and equilibration as they cannot be bypassed in one dimension. Interestingly, signatures of this anomalously slow dynamics in the thermal phase can be observed in the scaling of the optical conductivity~\cite{Agarwal,PhysRevB.92.104202}
\begin{equation}
\sigma(\omega) \sim \omega^{1-2/z},
\end{equation}
 or in the algebraic relaxation of (sum) of local observables~\cite{PVPtransition,PhysRevB.93.060201} such as the charge imbalance used in recent cold atom experiments~\cite{Schreiber842} to characterize the MBL phase. The general scaling structure of the phase transition is summarized in Fig.~\ref{fig:PhaseDiagramMBLETH}.

There remain many interesting questions regarding the nature of the MBL transition, and the current RG approaches should mostly be considered as a first step towards a complete understanding. Among the most challenging and controversial questions are the nature of the level statistics at the critical point~\cite{PhysRevB.93.041424,2015arXiv151008322M}, the existence of an intermediate phase that would be delocalized but non-ergodic~\cite{2014arXiv1405.1471G,PhysRevLett.113.046806,2015arXiv150103853P,2016arXiv160304701M}, the scaling of the eigenstate entanglement entropy in the critical fan~\cite{2014arXiv1405.1471G}, and the connection to experiments~\cite{Schreiber842,2015arXiv150807026S} that will hopefully help understanding the discrepancy between RG approaches and numerics. It would also be very interesting to understand the nature of the MBL transitions in higher dimensions, in driven (Floquet) systems, or driven by energy density (addressing the controversial issue of the existence of many-body mobility edges~\cite{PhysRevB.93.014203}).

\section{Conclusion}

In this review, we have described how the infinite number of conservation laws of integrable and many-body localized systems lead to new states far away from equilibrium that defy our physical intuition based on standard thermodynamics and hydrodynamics. For such systems, thermalization does not occur, Fourier's law does not hold, and quantum order and universal phase transitions can arise at finite energy density even in one dimensional systems, even if it is strictly forbidden by usual statistical mechanics arguments. For both types of systems, many open questions remain and we have given some specific examples in the final parts of Sections 3 and 4.

On the integrable side, it would be very interesting to develop a generalized hydrodynamical framework that would capture the structure of non-equilibrium steady-states in integrable systems, or to generalize the setup of Fig.~\ref{Fig1} to include, for example, a quantum impurity between the two reservoirs (see {\it e.g.}~\cite{PhysRevB.73.245326}).  The relation to steady states generated using open quantum systems and non-hermitian boundary conditions remains also mysterious.  There have been efforts in the high-energy community, often using holography, to obtain fundamental limits on hydrodynamics~\cite{kss} and thermalization, and also classifications of alternative hydrodynamics~\cite{rangamani}, but at the moment integrable models lie outside most such efforts and it would be nice to find connections, especially given the recent efforts to observe hydrodynamical behavior in electronic materials.

On the disordered side, the field of many-body localization is still in its infancy, and we expect the future to hold many interesting developments. Among some of the exciting issues, let us mention the role of symmetries on MBL phases and phase transitions, the nature of the MBL transition in one and higher dimensions, the classification of the various types of quantum order that can be stabilized at finite energy density using localization, or the interplay between quantum order, Floquet dynamics and many-body localization.  We expect that connections between MBL-type and conventional integrability will continue to emerge in the coming years, along with additional unexpected consequences of their failure to reach thermal equilibrium.

\paragraph{Acknowledgments.}

We thank J. Bardarson, F. Essler, Y. Huang, R. Ilan, M. Serbyn and especially C. Karrasch, S.A. Parameswaran and A.C. Potter for numerous discussions and collaborations on topics related to nonequilibrium dynamics in integrable and MBL systems. We also acknowledge support from the Department of Energy through the LDRD program of LBNL (R.V.), from NSF DMR-1507141 and a Simons Investigatorship (J.E.M.), and center support from CaIQuE and the Moore Foundation's EPiQS initiative.

\ \

\bibliography{MBL}

\end{document}